\documentclass[a4paper,11pt]{article}
\usepackage{jinstpub} 
\usepackage{lineno}
\usepackage{hyperref}
\usepackage{amsmath}
\usepackage{natbib}
\usepackage{booktabs}

\usepackage{float}
\usepackage{enumitem}

\usepackage{rotating}

\usepackage{caption}
\usepackage{subcaption}
\usepackage{graphicx}

\usepackage{xfrac}
\usepackage{xcolor}
\usepackage{scalerel}
\title{\boldmath Particle identification in the GlueX detector with machine learning}

\author[a, b]{\href{https://orcid.org/0009-0003-9547-0952}{E. Habjan}}
\author[a, c]{\href{https://orcid.org/0009-0003-8192-2313}{R. Dube}}
\author[a]{\href{https://orcid.org/0009-0007-9775-8158}{J. M\textsuperscript{\underline{c}}Intyre}}
\author[a]{M. Edo}
\author[a]{\href{https://orcid.org/0000-0002-1410-6012}{R. Jones}}
\affiliation[a]{University of Connecticut, Department of Physics,\\196 Auditorium Road, Unit 3046, Storrs, CT 06269, USA}
\affiliation[b]{Northeastern University, Department of Physics,\\100 Forsyth Street \#111, Boston, MA 02115, USA}
\affiliation[c]{Indiana University, Department of Physics,\\727 E. Third St., Swain Hall West Room 152, Bloomington, IN 47405}

\emailAdd{habjan.e@northeastern.edu}

\abstract{In particle physics experiments, identifying the types of particles registered in a detector is essential for the accurate reconstruction of particle collisions. At Thomas Jefferson National Accelerator Facility (Jefferson Lab), the GlueX experiment performs particle identification (PID) by setting specific thresholds, known as cuts, on the kinematic properties of tracks and showers obtained from detector hits. Our research aims to enhance this cut-based method by employing machine-learning algorithms based on multi-layer perceptrons and boosted decision trees. Similar approaches have been applied in other particle physics experiments and offer an opportunity to increase PID accuracies using reconstructed kinematic data. Our study illustrates that both multilayered perceptrons and boosted decision trees can identify charged and neutral particles in Monte Carlo simulated GlueX data with significantly improved accuracy over the current cuts-based PID method.}


\begin{document}
\maketitle
\flushbottom

\section{Introduction}
\label{sec:intro}
Since the 1970s, there has been significant interest in understanding the mechanism behind the confinement of quarks and gluons within quantum chromodynamics (QCD). The color confinement principle describes how quarks and gluons, which carry color charge, are permanently bound together to form protons, neutrons, and other color-neutral hadrons. This phenomenon prevents quarks and gluons from existing as free particles, ensuring they are always observed within larger, composite, color-neutral particles.

The constituent quark model (CQM)~\cite{Schematic_CQM} treats the simplest hadron configuration, a meson, as a bound quark-antiquark ($q\bar{q}$) state grouped into SU(N) flavor multiplets. While the CQM provides a natural framework to classify mesons, it is unable to explain the mechanism of quark confinement since it does not include the gluon, the mediator of the strong force, which also carries color charge in QCD. Hybrid mesons (e.g., $q\bar{q}g$ in a constituent gluon notation) can possess quantum numbers ($J^{PC}$) not possible from the CQM ($q\bar{q}$); therefore, the discovery of these \emph{exotic hybrid mesons} would be confirmation of physical states beyond the CQM. 

The GlueX experiment, located in Hall D at Jefferson Lab, searches for evidence of gluonic degrees of freedom in the excitation spectrum of light mesons through the mapping of the spectrum of mesons with exotic quantum numbers with masses up to 2.5 $\mathrm{GeV}/c^{2}$. Accurate identification of these mesons requires knowledge of their production mechanisms, quantum numbers ($J^{PC}$), and decay modes. This is accomplished through the partial wave analysis (PWA) of exclusive final states. To accomplish this, the GlueX experiment relies upon high statistics, linear polarization of the incident photon beam, excellent measurement resolution, full acceptance in decay angles, and correct decay particle identification (PID).

\begin{figure*}
    \centering
    \includegraphics[width=0.9\textwidth]{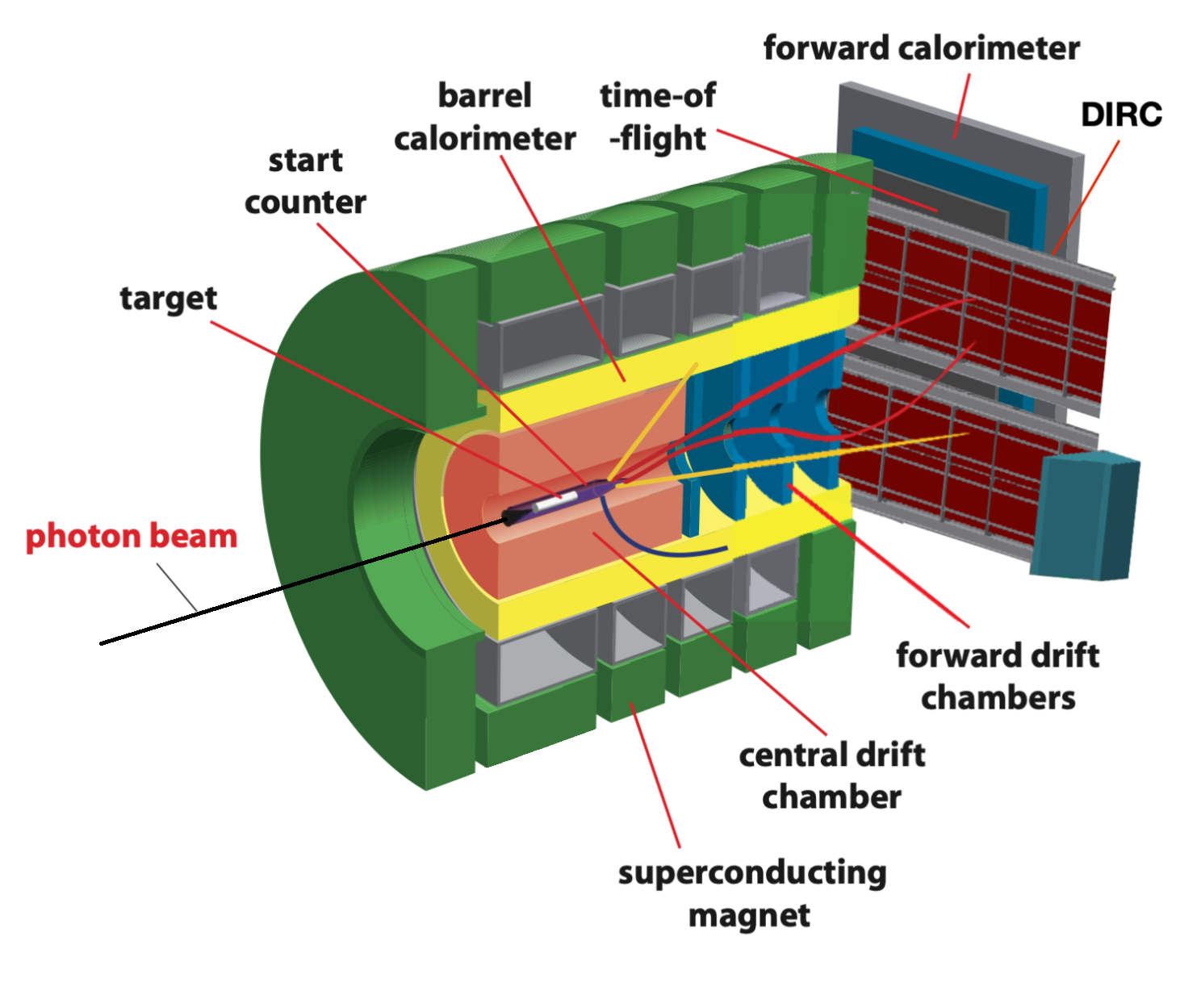}
    \caption{Cut-away view of the GlueX spectrometer.}
    \label{fig:GlueX}
\end{figure*}

Particle identification (PID) in the GlueX detector~\cite{GlueX2021}, depicted in Figure~\ref{fig:GlueX}, is accomplished using seven detector systems:
\begin{itemize}
    \item \textbf{SC:} a start counter consisting of a cylindrical array of 40 scintillator paddles immediately surrounding the liquid hydrogen target;
    \item \textbf{TOF:} a forward time-of-flight wall of 44 vertical and 44 horizontal scintillator paddles located in the forward direction downstream of the target;
    \item \textbf{BCAL:} a cylindrical geometry barrel calorimeter with alternating layers of lead and scintillating fiber located at large polar angles surrounding the target;
    \item \textbf{FCAL:} a forward lead-glass block calorimeter with a planar geometry orthogonal to the photon beam axis, located in the far-forward region downstream of the target;
    \item \textbf{CDC:} a central drift chamber detector consisting of layers (strung in axial and stereo configurations) of cathode straw tubes, each containing an anode wire and filled with a mixture of argon and CO$_{2}$ gas subtending polar angles greater than 20$^{\circ}$ surrounding the target;
    \item \textbf{FDC:} a forward drift chamber detector consisting of 24 planar drift chambers with cathode strip and wire readouts, and filled with a mixture of argon and CO$_{2}$ gas subtending polar angles less than 30$^{\circ}$ downstream of the target;
    \item \textbf{DIRC:} a Detection of Internally Reflected Cherenkov light detector used for forward region PID, located directly in front of the forward TOF.
\end{itemize}

The SC envelopes the target cell, covering $\sim$90\% of 4$\pi$ solid angle, and is the first detector to measure charged particles produced by interactions in the target. Due to its proximity to the target cell, the SC provides a timing signal relatively independent of particle type and trajectory and provides a fast signal that is used in the level-1 trigger for the experiment. The thin scintillator SC is used to identify the 4 ns accelerator electron radio-frequency (RF) beam bucket associated with detected particles. Energy deposition $(\mathrm{d}E/\mathrm{d}x)$ in the SC, in combination with the flight time from the TOF, is utilized for charged particle identification. The TOF provides pion and kaon separation up to a momentum of about 2 $\mathrm{GeV}/c$ and is located just upstream of the FCAL, 5.5 m downstream of the target, covering polar angle $\theta \in$ [1$^{\circ}$, 11$^{\circ}$]. The TOF provides PID through the measurement of a particle's velocity in the low momentum range and time of flight information with respect to the accelerator RF clock.

For charged particles emitted at angles greater than 10$^{\circ}$, the time of flight is measured in the BCAL. The BCAL is a barrel-shaped electromagnetic calorimeter comprised of a matrix of lead and scintillating fibers residing inside the GlueX 2.08 Tesla solenoid magnet, covering a polar angle $\theta \in$ [11$^{\circ}$, 126$^{\circ}$]. The BCAL measures the time and energy deposited by charged and neutral particles. The FCAL comprises 2800 lead-glass modules, each coupled to a photomultiplier (PMT) and stacked in a circular array inside a light-tight dark box enclosure. Decay particles that interact with the FCAL create an electromagnetic shower that is read out by the PMTs. 

The charged particle tracking chambers work together to provide a complete picture of the charged particle tracks, from low to high momentum, and from central to forward angles. The DIRC~\cite{DIRC_Detector} is located about 4 m from the target and covers a polar angle $\theta \in$ [2$^{\circ}$, 11$^{\circ}$]. As charged particles pass through the DIRC, they emit Cherenkov light while traveling inside the fused silica radiator bar. A fraction of this light is transported to the photon camera, where it is expanded and imaged on a pixelated photodetection plane. The resulting pattern on the pixelated photodetection plane provides information about the velocity of the charged particle, given the momentum vector reconstructed by the tracking system.

In this work, we aim to demonstrate the potential of ML~\cite{MLReview2021} as a tool to improve PID accuracies for GlueX analyses. ML has already been applied to PID in other high energy experiments \cite{Kasak_2024, Karwowska_2023, NA62_2023, Ryzhikov_2023, Graczykowski_2022} and has even been applied to a GlueX study~\cite{Barsotti2020} that used a neural network for classification to reduce background in the FCAL. Motivated by such studies and the demand for high-accuracy PID, this work makes use of a trained Multi-Level Perceptron (MLP) and Boosted Decision Tree (BDT) to classify single hadronic particles using GlueX MC simulation data. Cuts-based PID methods currently utilized by GlueX are carried out on the same MC simulation data sets and compared to the PID accuracies obtained from the charged particle MLP and BDT models.

\section{Monte Carlo Simulation Data Set}
\label{sec:data} 
The training and test data sets used for this study were extracted from low-momentum GlueX particle gun simulations (see Table~\ref{table:Particle_Gun}). In these simulations, a particle is spawned at a randomized location, known as a vertex, within the 30 cm long liquid H$_{2}$ target cell. The generated particle is launched from the assigned vertex in a random direction and a random momentum less than 1 $\mathrm{GeV}/c$. Detailed simulations of interactions between the generated particle and the GlueX detector and any subsequent decays and interactions are performed using the standard GlueX GEANT4-based software package~\cite{Geant4_2010, Geant4_2016}. The resulting simulated detector signals (``hits'') are then reconstructed using the \texttt{halld$\_$recon} package for shower and track identification. The kinematic parameters describing each of the reconstructed particles are saved for each event, together with the labels that identify the actual particle type and momentum that were fed as input to the simulation for each event. The complete list of features for each event that are provided to the models as input is shown in Table~\ref{tbl:dataset}.

\begin{table*}[htbp]
    \fontsize{10pt}{13.25pt}\selectfont
    \centering
    \caption{List of the particles generated using MC particle gun simulations for training, validating, and testing the MLP and BDT PID models. Note that each particle type is represented by 80,000 events in the training dataset, 20,000 events in the validation dataset, and 20,000 events in the test dataset for a total of 120,000 events.} 
    \smallskip
    \begin{tabular}{cccccc}
    \toprule
        Particle & Gen. Events & Undetected Rate & Multiple Particles Rate & Multiple Vertices Rate & Acceptance
        \\\midrule
        $p$  & 340436 & 0.637 & 0.011 & 0.000 & 0.352 \\
        $\bar{p}$  & 362205 & 0.295 & 0.374 & 0.000& 0.331 \\
        $K^{+}$  & 771870 &  0.057 & 0.011& 0.777 & 0.155\\
        $K^{-}$  & 336285 & 0.328 & 0.031 & 0.285 & 0.357 \\
        $\mu^{+}$  & 270888 & 0.375& 0.024 & 0.159& 0.443 \\
        $\mu^{-}$  & 218444 & 0.422 & 0.026 & 0.003& 0.549 \\
        $\pi^{+}$  & 299073 & 0.210& 0.022 & 0.366& 0.401 \\
        $\pi^{-}$  & 252078 & 0.366& 0.027 & 0.131& 0.476 \\
        $e^{+}$  & 230366 & 0.452 & 0.027 & 0.000& 0.521 \\
        $e^{-}$  & 229948 & 0.452 & 0.026 & 0.000& 0.522 \\[8pt]
        
        $\gamma$  & 171839 & 0.249 & 0.053 & 0.000 & 0.698 \\
        $K_{L}^{0}$  & 553815 & 0.422 & 0.108 & 0.253& 0.217 \\
        $n$ & 846048 & 0.844 & 0.014 & 0.000 & 0.142 \\
    \bottomrule
    \end{tabular}
    \small
    \label{table:Particle_Gun}
\end{table*}

To ensure that the tracks and showers that are recorded are directly associated with the generated particle, several event selection criteria were imposed. For example, some resulting particle decays and interactions observed within the detector produce tracks and showers that are not directly attributed to the generated particle. To reduce these types of events, events from the training and test samples are excluded where MC-generated particles decay before reaching either the BCAL or the FCAL. In Table \ref{table:Particle_Gun}, these events for which the particle decayed before producing a shower are labeled "multiple vertices" events.

To eliminate events where interactions with the detector produced secondary tracks or showers, limits (i.e., cuts) were placed on the number of tracks and showers per event in our data sets. Events generating a charged particle (e.g., $\mu^{\pm}$, $\pi^{\pm}$) were only included in the data sets if the reconstructed event contained only one track with a single track-associated calorimeter cluster. Meanwhile, events that generated neutral particles (e.g., $n$, $\gamma$, $K_{L}^{0}$) were required to have precisely one calorimeter cluster and no tracks. Events for which there are multiple tracks or showers are labeled "multiple particles" events in Table \ref{table:Particle_Gun}. Neutral particle events for which there are no showers, or charged particle events for which there are no tracks with an associated shower, are also excluded and are labeled "undetected" events in Table \ref{table:Particle_Gun}.

Although these cuts may inflate the accuracy of the manual and MLP PID technique due to the exclusion of complicated interactions with the detector, these cuts were necessary to ensure the event label matched the particle producing the shower or track in the training and test data sets. The results of this study can be interpreted as an upper bound for analogous methods applied to real data or simulation data with less strict selection criteria.

\newpage

In constructing the training and validation data set, our study included only the reconstructed track hypothesis that matched the generated particle type, which included $8 \times 10^{4}$ events per particle type. Since no muon hypothesis existed in the default reconstruction software, the charged pion track hypothesis was used in the training data set for events that generated a muon. Each row of the training data set represented a different event, while a row in the test and validation data sets corresponds to a PID hypothesis with 2-4 hypotheses per event. The test and validation data sets contain $2 \times 10^{4}$ events each per particle type. However, the number of rows in each data set is substantially larger due to the inclusion of multiple hypotheses per event.

\begin{table*}[htbp]
    \centering
    \fontsize{1pt}{12pt}\selectfont
    \small
    \caption{Reconstructed features of the particle gun data set. Note that the null values are only used for the MLP; BDTs have native support for missing values.}
    \smallskip
    \begin{tabular}{lclc}
        \toprule
        Column & Unit & Description & Null Value
        \\\midrule
        \texttt{E} & $\mathrm{GeV}$ & Particle total energy & -5\\
        \texttt{px} & $\mathrm{GeV}/c$ & Particle momentum X-component & -500\\
        \texttt{py} & $\mathrm{GeV}/c$ & Particle momentum Y-component & -500\\
        \texttt{pz} & $\mathrm{GeV}/c$ & Particle momentum Z-component & -500\\
        \texttt{q} & $e$ & Particle charge & -10\\ 
        \texttt{E1E9} &  & E1/E9 ratio for the matched FCAL cluster & -5\\
        \texttt{E9E25} &  & E9/E25 ratio for the matched FCAL cluster & -5\\
        \texttt{docaTrack} & cm & Impact parameter of track to FCAL cluster & -5\\
        \texttt{preshowerE} & GeV & Shower energy in the 1st layer of the BCAL & -5\\ 
        \texttt{sigLong} & cm & RMS of hit distance along the central axis of BCAL shower & -5\\ 
        \texttt{sigTrans} & cm & RMS of radial distance of hits from central axis of BCAL shower & -5\\
        \texttt{sigTheta} & rad & RMS of polar angle of hits around  central axis of BCAL shower & -5\\ 
        \texttt{E$\_$L2} & GeV & Shower energy in the 2nd layer of the BCAL & -5\\
        \texttt{E$\_$L3} & GeV & Shower energy in the 3rd layer of the BCAL & -5\\
        \texttt{E$\_$L4} & GeV & Shower energy in the 4th layer of the BCAL & -5\\
        \texttt{dEdxCDC} & $\mathrm{keV}/ \mathrm{cm}$ & Average $\mathrm{d}E/\mathrm{d}x$ of track in the CDC & -5\\ 
        \texttt{dEdxFDC} & $\mathrm{keV}/ \mathrm{cm}$ & Average $\mathrm{d}E/\mathrm{d}x$ of track in the FDC & -5\\
        \texttt{tShower} & ns & Mean shower time in the BCAL or FCAL & -10\\
        \texttt{thetac} & rad & Track Cerenkov angle measured by DIRC & -5\\
        \texttt{bCalPathLength} & cm & Track distance from vertex to BCAL entry & -5\\
        \texttt{fCalPathLength} & cm & Track distance from vertex to FCAL entry & -5\\
        \texttt{dEdxTOF} & $\mathrm{keV}/ \mathrm{cm}$ & Average track $\mathrm{d}E/\mathrm{d}x$ in the TOF & -5\\
        \texttt{tofTOF} & ns & Time from track vertex to impact on the TOF & -5\\ 
        \texttt{pathLengthTOF} & cm & Distance from track vertex to impact on the TOF & -5\\
        \texttt{dEdxSc} & $\mathrm{keV}/ \mathrm{cm}$ & $\mathrm{d}E/\mathrm{d}x$ of track in the SC & -5\\
        \texttt{pathLengthSc} & cm & Distance from track vertex to impact on the SC & -100\\ 
        \texttt{tofSc} & ns & Time from track vertex to impact on the SC & -100\\ 
        \texttt{xShower} & cm & Shower X-component & -500\\ 
        \texttt{yShower} & cm & Shower Y-component & -500\\ 
        \texttt{zShower} & cm & Shower Z-component & -500\\ 
        \texttt{xTrack} & cm & Track X-component & -500\\ 
        \texttt{yTrack} & cm & Track Y-component & -500\\ 
        \texttt{zTrack} & cm & Track Z-component & -500\\ 
        \texttt{CDChits} & & Number of straws in the CDC producing hits & -5\\
        \texttt{FDChits} & & Number of anode wires in the FDC producing hits & -5\\
        \texttt{DOCA} & cm & Impact parameter of track at the BCAL cluster & -5\\ 
        \texttt{deltaz} & cm & Impact parameter of track at the BCAL along Z & -100\\
        \texttt{deltaphi} & rad & Impact parameter of track at the BCAL along azimuth & -10\\
        \bottomrule
    \end{tabular}
    \small
    \label{tbl:dataset}
\end{table*}

\section{General Machine Learning Methodology} \label{sec:general ml methods}

In this Section, we explain and justify the implementations of the Cross-Entropy loss function,the Hyperband optimization, postprocessing steps, and SHAP values for the MLP and BDT models.

\subsection{Cross-Entropy Loss Function} \label{subsec:crossentropy}
The advent of logistic regression~\cite{Cox1958} and the creation of cross-entropy in the early years of information theory have evolved into a loss function ubiquitous in ML, referred to as the Cross-Entropy loss function. Minimizing cross-entropy between two distributions is equivalent to maximizing the log-likelihood~\cite{Shangnan2021}. 

\begin{gather}\begin{aligned}\label{eq:cross entropy func}
H(P,Q) = -\sum_{n = 1}^N Q_n \log{P_n}
\end{aligned}\raisetag{0\baselineskip}
\end{gather}

\noindent for a classifier with $N$ non-overlapping categories. In the case of this study, $N$ represents the number of particle types that the algorithm is trained to distinguish, and $P_n$ is the probability that the algorithm assigns to each particle type $n$ for a given event recorded in the detector, while $Q_n$ is 1 for $n$ corresponding to the label of the particle that produced the event, and zero for all others. Thus, the expression for $H(P,Q)$ reduces to the form, 

\begin{gather}\begin{aligned}\label{eq:cross entropy func l}
H(P,Q) = - \log{ P_{\ell} }
\end{aligned}\raisetag{0\baselineskip}
\end{gather}

\noindent where $\ell$ is the value of $n$ corresponding to the event label. The \texttt{Tensorflow} implementation of the cross-entropy loss function is used during training for the MLP models, while the BDT models used the \texttt{sklearn} implementation.

\subsection{Hyperband Algorithm}\label{subsec:hyperband}

To ensure that the model architectures are optimized for the particle identification problem, the \texttt{Hyperband} algorithm~\cite{Hyperband2016} is used to optimize the hyperparameters of the respective models.
 Hyperband is chosen as the optimization algorithm due to its computational efficiency and comparable performance compared to Bayesian optimization~\cite{Hyperband2016}. This algorithm selects different sets of hyperparameters and trains the models for a fixed number of epochs. It employs a technique known as \emph{successive halving}, where only half of the models with the lowest cross-entropy loss are allocated resources to continue training after a specified number of epochs have passed. This procedure is repeated until only a single set of hyperparameters remains, which are then used to train the respective models that are used for PID.
 
Due to the differing structures of MLPs and BDTs, the specific hyperparameters to be tuned by the Hyperband algorithm differ. In the case of the MLP, the Hyperband algorithm optimized the number of hidden layers, the number of neurons in each hidden layer, and the learning rate. Because the BDT architecture is far simpler than the MLP, the only hyperparameters tuned by Hyperband were the maximum tree depth and the learning rate. The varied hyperparamters of the MLPs and BDTs are shown in Tables~\ref{tbl:mlp_parameters} and~\ref{tbl:bdt_parameters}, respectively. The final optimized parameters of each model are discussed in Section \ref{subsec:hyperband results}.

\subsection{Postprocessing}
\label{sec:matching}

As described in Section \ref{sec:data}, the test and validation datasets contain one entry for each PID hypothesis that was successfully used to reconstruct a track. This design avoids bias introduced by selecting only the entry whose hypothesis matches the true particle type (as was done for the training data), though it necessitates postprocessing of the model predictions to ensure that only one prediction is made per track, rather than one per hypothesis.

For each track, each model produces a prediction for each available hypothesis. However, only predictions that correspond to the hypothesis used in the reconstruction of the entry are considered valid, since the model was trained exclusively on examples reconstructed under the correct hypothesis. If multiple valid predictions exist for a single track, the prediction with the highest confidence score is selected. If no valid predictions exist for a track, the model returns "No ID". Note that this postprocessing only applies for the charged particles, as hypotheses are not used in the standard reconstruction of showers in the GlueX detector.

\subsection{SHAP Values}
\label{subsec:shap values methods}

To analyze the PIDs made by the BDT and MLP models, \textbf{Sh}apley \textbf{A}dditive ex\textbf{P}lanations (SHAP)~\citep{Lundberg2017} is employed to assess the importance of each feature used in training. SHAP is derived from Shapley values from cooperative game theory and is a method to measure the average contribution of a given feature across the entire feature space. A SHAP value is computed for each feature for a given classifier by considering possible permutations of features and then taking the average of all marginal contributions by a feature to the resultant prediction. Because this process can be extremely computationally expensive, 1000 hypotheses from each generated particle type are randomly selected for use in SHAP calculations. The background events are utilized differently by the two models. The \texttt{DeepExplainer} from the \texttt{shap} library (used for the MLP) randomly selects 1000 background events per particle type from the training data. The \texttt{TreeExplainer} from the \texttt{shap} library (used for the BDT) is able to use the distribution of training samples in each leaf of the tree, meaning the \texttt{TreeExplainer} uses the entire training sample for its SHAP calculations \cite{treeexplainer}. This difference in treatment is due to the fact that the \texttt{DeepExplainer} object integrates over the entire background sample for each sample, which can be extremely inefficient if the entire training sample is used for the background distribution \cite{deepexplainer}.

\section{PID Methods}
\label{sec:PID Methods} 

The methodology of the cuts-based, MLP and BDT PID are discussed in this section. The cuts of the standard method PID is optimized using the training dataset while the MLP and BDT models are trained with the training dataset and use the validation dataset to achieve optimized model hyperparameters.

\subsection{Cuts-Based PID} \label{subsec:cuts-based pid}

In this work, we identify pions ($\pi^{\pm} \approx 140 \, \sfrac{MeV}{c^{2}}$) and muons ($\mu^{\pm} \approx 106 \, \sfrac{MeV}{c^{2}}$) as the same particle class, denoted as $\pi^{+} \lvert \, \mu^{+}$ or $\pi^{-} \lvert \, \mu^{-}$. Our PID method introduces this simplification since pions and muons have similar masses, and the GlueX detector does not have a hadronic calorimeter, which makes distinguishing between these particles difficult. Muons can be distinguished from pions by the shapes of the clusters they leave in the FCAL. However, this must be performed during event reconstruction and is outside the scope of this paper.

The timing cuts implemented in this paper are taken from the GlueX standard PID for Spring 2017-2019 data ~\cite{GlueX2021} and each is shown in Table~\ref{tbl:manualPID}. The measured BCAL and FCAL times were recorded as a single variable, \texttt{tShower}, in the simulation data set; thus, the BCAL and FCAL time measurements must be distinguished. If an event has a detection for \texttt{E$\_$L2}, then \texttt{tShower} is the mean shower time in the BCAL, and if there is a detection for \texttt{E1E9}, then \texttt{tShower} is the mean shower time in the FCAL. The mean shower times in each detector are compared with the calculated time from the vertex to the BCAL (\texttt{tFlightBCAL}) or the FCAL (\texttt{tFlightFCAL}). A $\chi^{2}$ value is calculated between the mean shower time and the calculated shower times. Only hypotheses with a $\chi^{2}$ value of less than 0.075 are considered; any hypotheses above this threshold are labeled as no identification (no ID). A logarithmic grid search of different $\chi^{2}$ values in the range [$10^{-2}$, $10^{0}$] was conducted using the training dataset in order to optimize the training accuracy of the timing cuts. Only timing information is available for charged particles in the simulation data set; therefore, no timing cuts were made for neutral particles. 

In addition to timing cuts, track energy loss cuts are implemented using the \texttt{dEdxCDC} variable and the magnitude of the particle momentum. To create a decision boundary between each particle, the functional form of the equations utilized in the GlueX standard PID for Spring 2017-2019 data is used, as shown in Equation \ref{eq:deds_eq}.  

\begin{gather}\begin{aligned}\label{eq:deds_eq}
f_{\scaleto{i\mathstrut}{7pt}}(p) = \exp(a_{\scaleto{i\mathstrut}{7pt}}p + b_{\scaleto{i\mathstrut}{7pt}}) + c_{\scaleto{i\mathstrut}{7pt}}
\end{aligned}\raisetag{0\baselineskip}
\end{gather}

\noindent where $p$ is the momentum ($\mathrm{GeV}/c$) of the reconstructed track in the CDC and the subscript $i$ denotes the constants and decision boundary that distinguishes two particle species. Using the training data set, the number of incorrectly identified particles is minimized by treating constants $a_{\scaleto{i\mathstrut}{7pt}}$, $b_{\scaleto{i\mathstrut}{7pt}}$, and $c_{\scaleto{i\mathstrut}{7pt}}$ as free parameters and utilizing the \texttt{minimize} method from the \texttt{scipy.optimize}~\cite{scipy} module. The energy deposition $(\mathrm{d}E/\mathrm{d}x)$ vs $p$ decision boundaries are only relevant for charged particles. The constants for the optimized energy deposition vs $p$ boundaries are shown in Table~\ref{table:parameters} 

\begin{table*}[htbp]
    \fontsize{10pt}{13.25pt}\selectfont
    \centering
    \caption{The optimized energy deposition $(\mathrm{d}E/\mathrm{d}x)$ vs $p$ boundary parameters where $f_3$ separates electrons from pions/muons, $f_2$ separates pions/muons/electrons from kaons, and $f_1$ separates kaons from protons/antiprotons.}
    \smallskip
    \begin{tabular}{ccccccc}
    \toprule
        $f_{\scaleto{i\mathstrut}{7pt}}(p)$ & \hspace{20pt} & $a_{\scaleto{i\mathstrut}{7pt}}$ [$\mathrm{GeV}/c$] & & $b_{\scaleto{i\mathstrut}{7pt}}$ & & $c_{\scaleto{i\mathstrut}{7pt}}$ [$\mathrm{keV}/ \mathrm{cm}$]
        \\\midrule
        $f_{\scaleto{1\mathstrut}{7pt}}(p)$ & & -5.095 & & -10.205 & & $2.080 \times 10^{-6}$ \\
        $f_{\scaleto{2\mathstrut}{7pt}}(p)$ & & -3.947 & & -12.284 & & $1.936 \times 10^{-6}$ \\
        $f_{\scaleto{3\mathstrut}{7pt}}(p)$ & & -0.185 & & -19.215 & & $2.190 \times 10^{-6}$ \\
    \bottomrule
    \end{tabular}
    \small
    \label{table:parameters}
\end{table*}
\begin{figure}
\centering
\includegraphics[width=\textwidth]{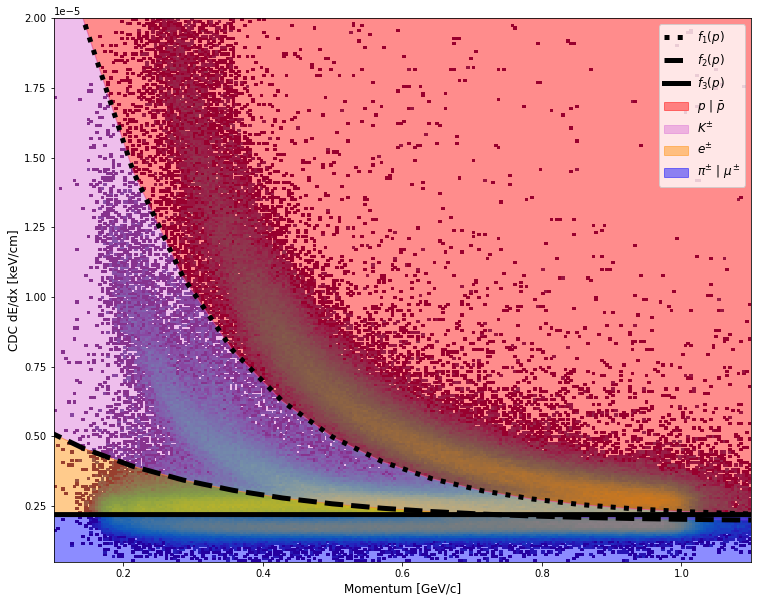}
\caption{A 2-dimensional histogram of the average track energy loss in the CDC plotted against the magnitude of momentum vectors from the test data set. The cuts-based PID parameters described in Section~\ref{subsec:cuts-based pid} are overlaid to show the classification boundaries; $f_{\scaleto{1\mathstrut}{7pt}}(p)$ is shown with the dotted line, $f_{\scaleto{2\mathstrut}{7pt}}(p)$ is shown with the dashed line, and $f_{\scaleto{3\mathstrut}{7pt}}(p)$ is shown with the solid line. The functional form of each decision boundary is shown in equation~\ref{eq:deds_eq} and the constants for each equation in Table~\ref{table:parameters}. Regions of the plot shaded in red are classified as $p$ or $\bar{p}$, purple as $K^{\pm}$, yellow as $e^{\pm}$, and blue as $\pi^{\pm}$ or $\mu^{\pm}$.}
\label{fig:PID-cuts}
\end{figure}

\begin{table*}[htbp]
    \fontsize{10pt}{13.25pt}\selectfont
    \centering
    \caption{Cuts-based PID parameters. If an entry is missing, there is no cut for that particle. The $f_1$, $f_2$ and $f_3$ cut curves corresponded to equation~\ref{eq:deds_eq} with variables listed in Table~\ref{table:parameters}.}
    \smallskip
    \begin{tabular}{ccccc}
        \toprule
        Particle & $\Delta t$ BCAL [ns] & $\Delta t$ FCAL [ns] & $\mathrm{d}E/\mathrm{d}x$ [$\mathrm{keV}/ \mathrm{cm}$] & \hspace{5pt} $E / p$ [$c$]
        \\\midrule
        $e^{\pm}$ & $\pm$ 1.0 & $\pm$ 2.0 &  $f_{\scaleto{3\mathstrut}{7pt}}(p)$ $<$ $\mathrm{d}E/\mathrm{d}x$ $<$ $f_{\scaleto{2\mathstrut}{7pt}}(p)$ & \hspace{5pt} > 0.83 \\[6pt]
        $\pi^{\pm} \lvert \, \mu^{\pm}$ & $\pm$ 1.0 & $\pm$ 2.0 & $\mathrm{d}E/\mathrm{d}x$ $<$ $f_{\scaleto{3\mathstrut}{7pt}}(p)$ & \hspace{5pt} < 0.83 \\[6pt]
        $K^{\pm}$ & \hspace{5pt}$\pm$ 0.75 & $\pm$ 2.5 & $f_{\scaleto{2\mathstrut}{7pt}}(p)$ $<$ $\mathrm{d}E/\mathrm{d}x$ $<$ $f_{\scaleto{1\mathstrut}{7pt}}(p)$ & \\[6pt]
        \hspace{-7pt}$p \, \lvert \, \bar{p}$ & $\pm$ 1.0 & $\pm$ 2.0 & $f_{\scaleto{1\mathstrut}{7pt}}(p)$ $<$ $\mathrm{d}E/\mathrm{d}x$ & \\[6pt]
        \bottomrule
    \end{tabular}
\small
\label{tbl:manualPID}
\end{table*}

Each optimized decision boundary is shown overlaid on the test data set in Figure~\ref{fig:PID-cuts}. An additional manual cut is made for electrons and muons/pions, in which the ratio of the particle's total energy \texttt{E} to the magnitude of the momentum is taken, with a decision boundary at $E / p = 0.83\,c$. A comprehensive summary of the cuts-based PID parameters for charged particles is shown in Table \ref{tbl:manualPID}. A PID is made for every hypothesis in our test data set that meets the $\chi^{2}$ criteria; however, if a given event met none of the criteria, then no ID is designated. Furthermore, if an event has two or more PIDs that match different hypotheses in the test data set, the particle type with the highest $\chi^{2}$ value is designated.

\subsection{Multi-Layer Perceptrons}
\label{subsec:mlp description}

With a sufficient number of parameters, MLPs can be used to approximate a map from one distribution to another. In the case of this work, we aim to approximate the relationship between the labels shown in Table \ref{tbl:dataset} and a particle type. During training, an optimizer adjusts the output distribution of an MLP to more closely match the target distribution. The MLPs that are trained in this study use the Adam optimizer~\cite{Adam2014} is used to optimize the objective function, i.e. classification accuracy. By combining the strengths of the AdaGrad~\cite{Adagrad_paper} and RMSProp~\cite{RMSprop_Lecture} optimization methods, Adam excels in first-order gradient-based optimization of stochastic functions. It uses estimates of the first and second moments of gradients to compute adaptive learning rates for each parameter, making it highly effective in high-dimensional parameter spaces. Due to the high level of stochasticity in experimental particle physics data, Adam is particularly well-suited for minimizing the Cross-Entropy Loss Function.

An MLP structure contains an input layer, one or more hidden layers, and an output layer. The input layer of our models is comprised of 38 nodes, which is equal to the number of feature labels used in training, shown in Table~\ref{tbl:dataset}. Each of the hidden layers of the MLP makes use of the \textbf{Re}ctified \textbf{L}inear \textbf{U}nit (ReLU) activation function~\cite{Hahnloser2000, Agarap2018}, and is defined in Equation~\ref{eq:relu}.

\begin{gather}\begin{aligned}\label{eq:relu}
f(x) = max(0 , x)
\end{aligned}\raisetag{0\baselineskip}
\end{gather}

\noindent For any input $x$ from a previous neuron, a non-negative output $f(x)$ will be produced from that neuron. The non-linearity of ReLU introduces sparsity and avoids saturation at large values while remaining simple. These advantages allow for computational efficiency during training and for meaningful connections to be drawn between complex relationships in the data. In the output layer of the MLP models, the softmax activation function is used and is defined by Equation~\ref{eq:softmax}.

\begin{gather}\begin{aligned}\label{eq:softmax}
\sigma(\mathbf{z})_i = \frac{e^{z_i}}{\sum_{j = 1}^{K} e^{z_j}}
\end{aligned}\raisetag{0\baselineskip}
\end{gather}

\noindent The softmax activation function computes a probability distribution for the outputs $z$ of an MLP. In Equation~\ref{eq:softmax}, the probability $\sigma(\mathbf{z})_i$ is computed for an input $z_i$. This insures that the sum of the probabilities will be $1$, allowing for a confidence-based prediction when classifying particles, as each probability correspond to each particle in the respective data set.

Since MLPs do not have native support for missing values, the training data for the MLP is first preprocessed to fill in missing features with the corresponding ``Null Value'' specified in Table~\ref{tbl:dataset}. The MLP models are then trained using the \texttt{TensorFlow} implementations of the Adam Optimizer and ReLU activation function described earlier in this section. The number of neurons, the number of hidden layers, and the learning rate of the Adam optimizer are optimized by \texttt{Hyperband} to have the maximum validation accuracy and are varied in the ranges shown in Table \ref{tbl:mlp_parameters}. Validation accuracy is calculated such that only one prediction is made for each event following the methodology presented in Section \ref{sec:matching}. If a prediction surpasses a threshold of $\sigma(\mathbf{z})_i > 0.4$, then a PID is made; below this threshold a no ID is assigned for that event. This threshold is chosen because the probability distributions produced from the softmax activation function are bimodal, and removing any classifications with $\sigma(\mathbf{z})_i < 0.4$ yield the highest purities. The optimized hyperparameters are used to train an MLP model for up to 50 epochs. The training ends early if the event-based validation accuracy changes by less than 0.001 after five successive epochs. 

\begin{table*}[htbp]
    \fontsize{10pt}{13.25pt}\selectfont
    \centering
    \caption{Variation for each hyperparameter permitted during training of the MLP.}
    \smallskip
    \begin{tabular}{lcc}
        \toprule
        \textbf{Hyperparameter} & \hspace{10pt} & \textbf{Variation Permitted}
        \\\midrule
        Hidden layers & & 1 -- 3 \\[6pt]
        Neurons per hidden layers & & 25 -- 400 \\[6pt]
        Learning rate & & $10^{-4}$ -- $10^{-2}$ \\
        \bottomrule
    \end{tabular}
\small
\label{tbl:mlp_parameters}
\end{table*}

\subsection{Boosted Decision Trees}

Decision trees are models that use sequential splits in feature space to sort data into categories or make predictions. Boosted Decision Trees models extend this idea by sequentially training trees, where each new tree is constructed to correct the errors of previous trees. The \texttt{xgboost} library, which is used in this study, achieves this via gradient boosting, in which subsequent trees are trained to minimize a loss function approximated through its gradient and Hessian\cite{general-xgboost}. The construction of individual trees is done using the approximate method provided by the \texttt{xgboost} library. This method is preferred because it is significantly faster than the exact greedy method while maintaining split accuracy superior to that of the default histogram-based method\cite{xgboost-methods}. Furthermore, unlike the exact greedy method, the approximate method allows for distributed training and GPU acceleration, which further decreases the training time.

Although the architecture of a boosted decision tree is not as flexible as that of a MLP, we are able to control the total number of trees in the final model, the maximum depth of each tree (the maximum number of branches along any path), and the shrinkage parameter (the gradient boosting equivalent of a learning rate).

\begin{table*}[htbp]
    \fontsize{10pt}{13.25pt}\selectfont
    \centering
    \caption{Variation for each hyperparameter permitted during training of the boosted decision trees model.}
    \smallskip
    \begin{tabular}{lcc}
        \toprule
        \textbf{Hyperparameter} & \hspace{10pt} & \textbf{Variation Permitted}
        \\\midrule
        Number of Boosting Rounds & & <500 \\[6pt]
        Max Depth & & 3 -- 15 \\[6pt]
        Learning rate & & $10^{-3}$ -- $0.3$ \\
        \bottomrule
    \end{tabular}
\small
\label{tbl:bdt_parameters}
\end{table*}

\section{Results and Discussion} \label{sec:results}

This Section presents the results of the traditional PID cuts, the MLP model and the BDT model. It directly compares these three methods and discusses the advantages of ML for post-reconstructed PID on the test dataset. The importance of each feature in the simulation data set is examined to better understand how the ML models make predictions. 

\subsection{Hyperband results}
\label{subsec:hyperband results}

As discussed in Section \ref{subsec:hyperband}, the parameters for both the MLP and BDT models were varied using the \texttt{HyperBand} algorithm to maximize the validation accuracy of each model that is trained. The optimal MLP for charged PID had two hidden layers and a total of $\sim$7.5 $\times$ 10$^{3}$ parameters. The first hidden layer of the charged MLP has 111 neurons, the second hidden layer has 27 neurons and the model was trained with a learning rate of $\sim1 \times 10^{-3}$. In comparison, the neutral MLP model that had only one hidden layer with $\sim$9.1 $\times$ 10$^{3}$ parameters. This hidden layer has 479 neurons and the model was trained with a learning rate of  $\sim2 \times 10^{-4}$. The optimal BDT for charged PID had a max depth of 15 with a learning rate of $\sim0.2$, while the optimal BDT for neutral PID had a max depth of 14 and a learning rate of $\sim0.17$.

\subsection{Charged PID Accuracies}

\begin{figure}[h]
    \centering
    \includegraphics[width=0.6\textwidth]{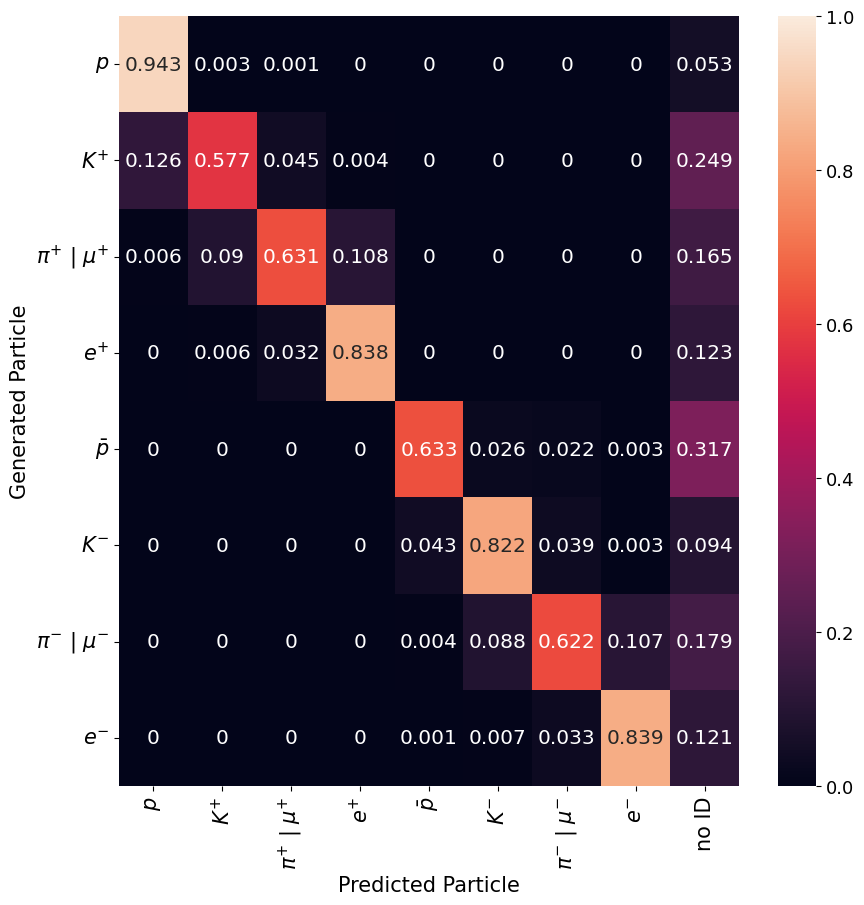}
    \caption{\label{fig:manual_PID}The confusion matrix for cuts-based PID on charged particles is shown. The generated particle is shown on the y-axis, and the identified particle is shown on the x-axis. A no ID classification was given for events in the cuts-based PID scheme that do not meet the $\chi^{2}$ criteria described in Section~\ref{subsec:cuts-based pid}.}
\end{figure}

\begin{figure}[htbp]
    \centering
    \begin{subfigure}[t]{0.6\textwidth}
        \centering
        \includegraphics[width=\textwidth]{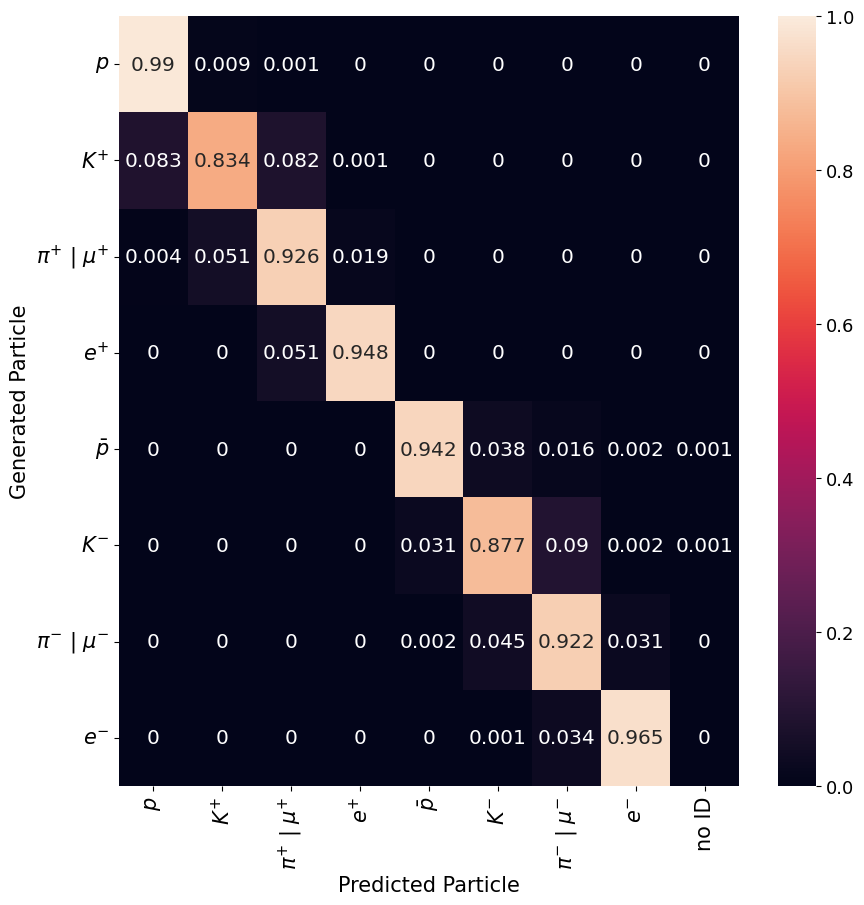}
        \caption{Charged MLP}
        \label{fig:mlp_charged_cm}
    \end{subfigure}
    \hspace{-0.7em}
    \begin{subfigure}[t]{0.4\textwidth}
        \centering
        \vspace{-21em}
        \includegraphics[width=\textwidth]{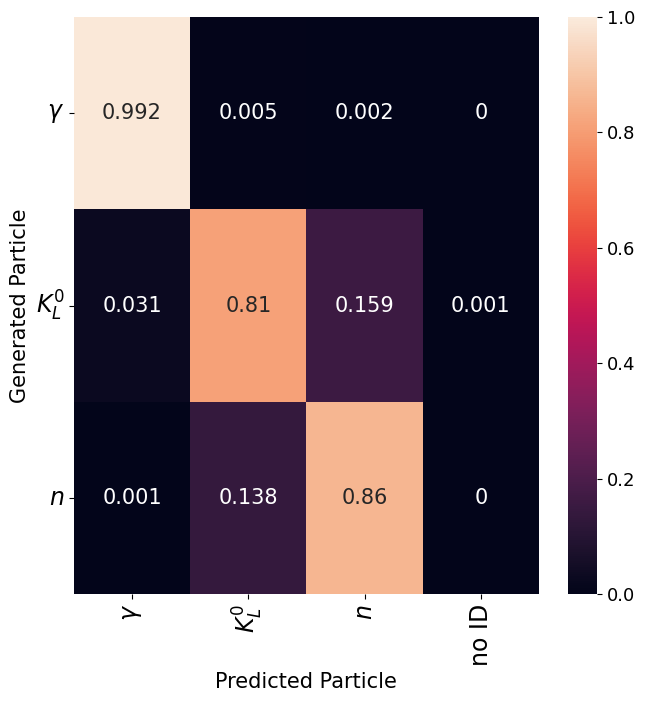}
        \caption{Neutral MLP}
        \label{fig:mlp_neutral_cm}
    \end{subfigure}
    \caption{The confusion matrix for MLP PID on charged particles is shown in Figure~\ref{fig:mlp_charged_cm} and the confusion matrix for MLP PID on neutral particles is shown in Figure~\ref{fig:mlp_neutral_cm}. The generated particle is shown on the y-axis, and the identified particle is shown on the x-axis. A no ID classification is given when the confidence criteria described in Section~\ref{subsec:mlp description} are not achieved.}
    \label{fig:mlp_pid}
\end{figure}

\begin{figure}[htbp]
    \centering
    \begin{subfigure}[t]{0.6\textwidth}
        \centering
        \includegraphics[width=\textwidth]{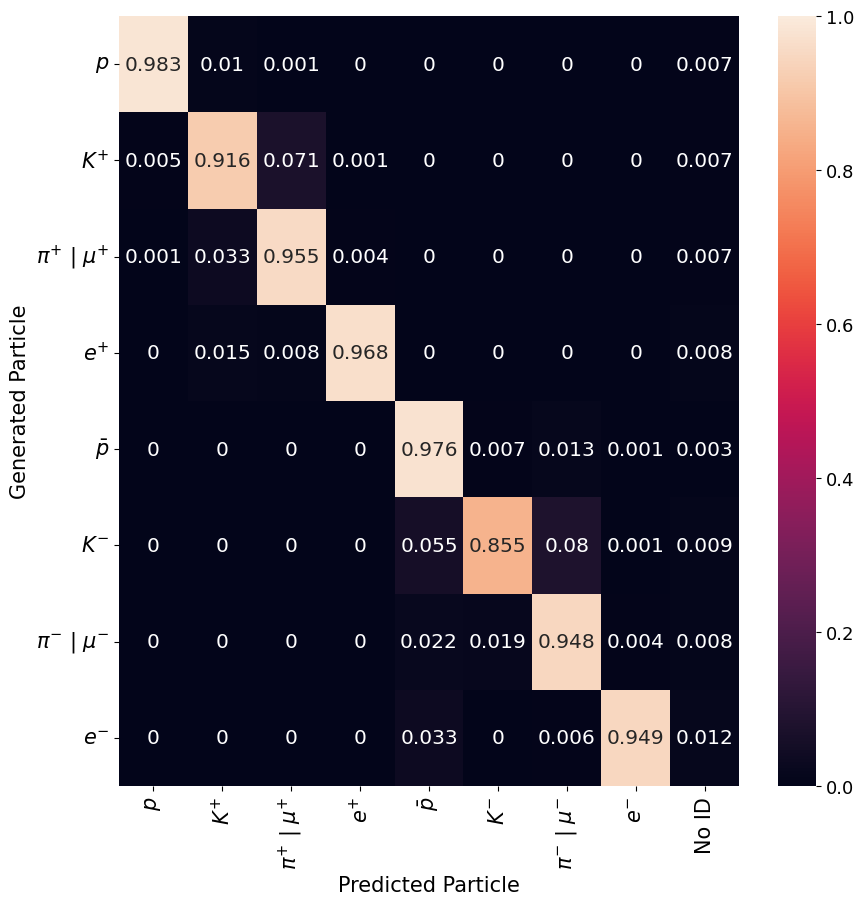}
        \caption{Charged BDT}
        \label{fig:bdt_charged_cm}
    \end{subfigure}
    \hspace{-0.7em}
    \begin{subfigure}[t]{0.4\textwidth}
        \centering
        \vspace{-21em}
        \includegraphics[width=\textwidth]{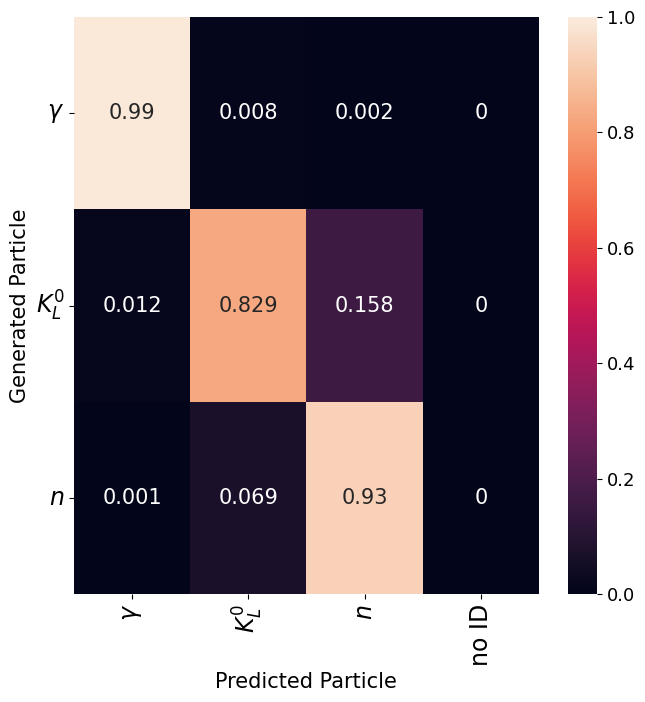}
        \caption{Neutral BDT}
        \label{fig:bdt_neutral_cm}
    \end{subfigure}
    \caption{The confusion matrix for BDT PID on charged particles is shown in Figure~\ref{fig:bdt_charged_cm} and the confusion matrix for BDT PID on neutral particles is shown in Figure~\ref{fig:bdt_neutral_cm}. The generated particle is shown on the y-axis, and the identified particle is shown on the x-axis.}
    \label{fig:bdt_pid}
\end{figure}

To assess the classification power of each PID method, we present the confusion matrices for the three PID methods described in Section~\ref{sec:PID Methods} evaluated on the test data sets described in Section~\ref{sec:data}. The confusion matrices for manual, MLP and BDT PID shown in Figures ~\ref{fig:manual_PID}, ~\ref{fig:mlp_pid}~and~\ref{fig:bdt_pid}, respectively, make use of the postprocessing described in Section~\ref{sec:matching}.

As shown in Figure \ref{fig:manual_PID}, the highest accuracy obtained using manual cuts is 94.3\% for the proton ($p$) sample, with only 5.3\% of the $4 \times 10^{4}$ simulated proton test events producing a no ID result. The $K^{-}$, $e^{-}$, and $e^{+}$ had cuts-based PID accuracies ranging from 82\% to 84\%. The poorest performance of the cuts-based PID occurred in $\bar{p}$, $K^{+}$, $\pi^{+} \lvert \, \mu^{+}$, and $\pi^{-} \lvert \, \mu^{-}$, which produced accuracies between 57\% and 64\%. Many of the samples that are incorrectly identified by the cuts-based PID method are designated as no ID. The most prominent exceptions are the $K^{+}$ test sample, which is misidentified as a $p$ 12.6\% of the time, and the $\pi^{+} \lvert \, \mu^{+}$ test sample that is misclassified as a $K^{+}$ or $e^{+}$ 9.0\% and 10.8\% of the time, respectively. The $\pi^{-} \lvert \,\mu^{-}$ test sample is misidentified 10.7\% of the time as an $e^{-}$ and 8.8\% as a $K^{-}$. Additionally, there are several cases in which the cuts-based PID method misidentified test sample particles less than 5\% of the time. The large number of events that did not pass the $\chi^{2}$ timing cut, along with the instances of substantial particle misidentification shown in Figure~\ref{fig:manual_PID}, underscore areas where ML-based PID methods can improve GlueX PID performance relative to cuts-based PID methods.

Figure~\ref{fig:mlp_charged_cm} shows the performance of our MLP PID method on charged particle MC simulation data displayed as a confusion matrix. Substantial improvements are achieved in particle identification for all charged particle MC data samples. The most notable improvement is the reduction of events classified as no ID. The only charged particles that have no ID classifications are $\bar{p}$ and $K^{-}$, however there are only $0.1 \%$ of events with the no ID label. This is a significant reduction from the manual PID method. The $\bar{p}$, $K^{+}$, $\pi^{+} \lvert \, \mu^{+}$, and $\pi^{-} \lvert \, \mu^{-}$ which had the lowest cuts-based PID accuracies in the range 57\% -- 64\%, but are correctly identified with accuracies in the range 83\% to 95\% by the MLP models. The $e^{+}$ and $e^{-}$ samples are both identified with 83.8\% accuracy by the cuts-based PID method, but are correctly identified by the MLP PID method with accuracies of 94.8\% and 96.5\%, respectively. Compared to the cuts-based PID method, the PID accuracy for all charged particles improved when the MLP PID method was implemented. Even the highest PID accuracy in the cuts-based PID method sample ($p$) is improved by 3.4\% by the charged MLP model. These results show that MLPs are better at PID on MC data, and that MLPs have the potential to significantly increase the PID accuracies for the GlueX experiment. 

Although PID performance increased for every class of charge particle using the MLP PID method, there are three instances where particle misclassifications increased. For the $\bar{p}$, the PID accuracy increased by 30.9\% and the percentage of no ID events decreased by 31.6\% compared to the cuts-based PID method, however there is a 1.2\% increase in misidentification as a $K^{-}$ for the MLP PID method. Similarly, the $K^{+}$ sample has a 3.7\% increase in being identified as a $\pi^{+} \lvert \, \mu^{+}$ when using the MLP PID method, but also correctly identified 25.7\% more events and reduced events with a no ID classification by 24.9\% when compared to cuts-based PID. The most significant increase in misidentification for the charged MLP model occurred for the $K^{-}$ sample in which there is a 5.1\% increase in misidentification as a $\pi^{-} \lvert \, \mu^{-}$, however, the  $K^{-}$ sample has a 5.5\% increase in correctly identified events by the MLP PID method and a 9.4\% reduction in no ID classification. In summation, the only downside of the charged MLP PID method is an increase in misclassifications for three charged particles in three particular cases. These increases can potentially be removed by using more training data and more complex ML architectures.

The confusion matrix for charged BDT PID is shown in Figure ~\ref{fig:bdt_charged_cm}. Overall, the charged BDT model outperforms the MLP with the exception of the $p$, $K^{-}$ and $e^{-}$ in which the MLP has on average $\sim1.5\%$ better classification accuracies. For the remaining 5 particles, the charged BDT model has better classification accuracies by $\sim3.8\%$. The largest increase in charged PID accuracy for the BDT method is the $K^{+}$, which is $8.2 \%$ better than the MLP method and $33.9 \%$ better than the manual PID method. Consequently, the misclassification of $K^{+}$ as $p$ and $\pi^{+} \lvert \, \mu^{+}$ decrease by $7.8 \%$ and $1.1 \%$, respectively. Furthermore, although the BDT PID method has lower accuracies for $p$, $K^{-}$ and $e^{-}$ compared to the MLP model, the BDT is more robust in regard to assigning misclassified particles to the no ID label when compared to the MLP. For example, the MLP has $2.2 \%$ better $K^{-}$ accuracy, but the BDT has a $K^{-}$ purity of $~97.0\%$ while the MLP has a $\sim91.3\%$ purity. In total, the charged BDT model has an average purity of $\sim95.3\%$ while the MLP has an average purity of $\sim92.7\%$. The average increase in accuracies as well as purity for the BDT compared to the MLP indicate that the BDT is superior in terms of PID on recontructed MC data from the GlueX detector.

\subsection{Neutral PID Accuracies}

Figure~\ref{fig:mlp_neutral_cm} shows the confusion matrix for the neutral particle MLP PID model. Unlike charged particles, robust cuts-based PID methods for classifying neutral particles do not exist in the GlueX experiment. The lone exception is the existence of timing cuts for $\gamma$. Unfortunately, the simulation data set did not recover any predicted timing values (e.g., \texttt{tFlightBCAL} and \texttt{tFlightFCAL}); thus, no cuts-based PID could be carried out for neutral particles. Regardless, the MLP and BDT PID method showed that neutral particles can be identified accurately. For the MLP, an accuracy of 99.2\% is achieved for identifying $\gamma$, 81.0\% for $K^{0}_{L}$, and 86.0\% for $n$. The BDT has a small decrease in $\gamma$ accuracy by $0.2\%$, but increases $K^{0}_{L}$, and $n$ classification accuracies by 2.1\% and 7.0\%, respectively. Similar to the charged particle models, the BDT, on average, outperforms the MLP in neutral classification. Interestingly, the large increase in $n$ accuracy primarily comes from a 6.9\% decrease in misclassification of $n$ as $K^{0}_{L}$, while the misclassification of $K^{0}_{L}$ as $n$ remains nearly the same (0.2~\% difference). These results show that the BDT outperforms the MLP models in both charged and neutral PID for reconstructed simulation data from the GlueX detector.

\subsection{SHAP Values}
\label{subsec:shap value results}

\begin{figure}[p]
    \centering
    \includegraphics[width=\textwidth]{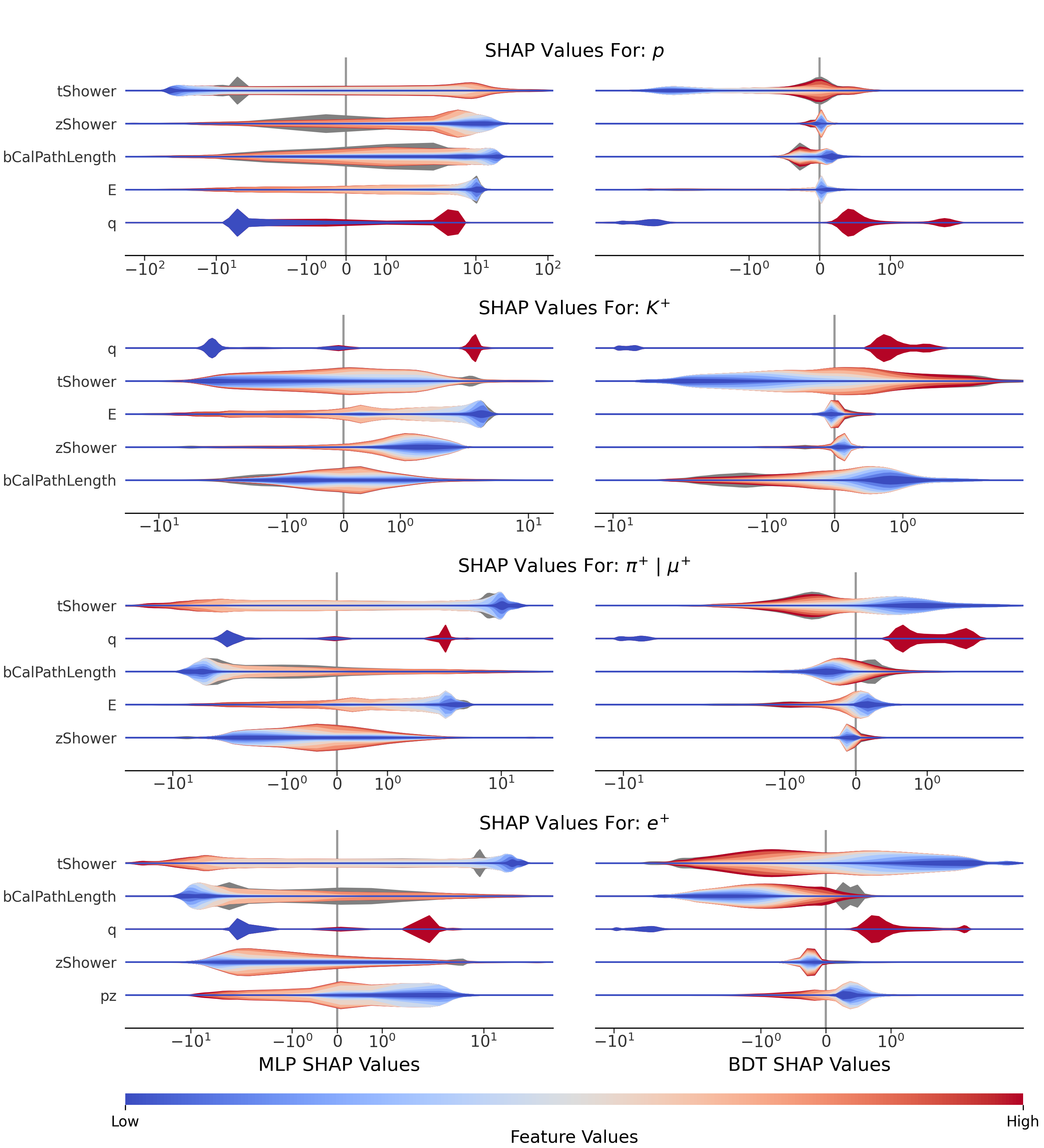}
    \caption{Layered violin plots showing SHAP value distributions for positively charged particles. For each class, the left plot corresponds to the multilayer perceptron (MLP) classifier and the right to the boosted decision tree (BDT). Each row represents an input feature, with the horizontal axis indicating SHAP values on a symmetric logarithmic scale. The width of each violin reflects the density of SHAP values, and the internal color gradient shows the corresponding feature value, with blue representing low values and red representing high values. Gray corresponds to missing values. Features are ordered by mean absolute SHAP value averaged between both models.}
    \label{fig:positive-charged-shap}
\end{figure}

\begin{figure}[p]
    \centering
    \includegraphics[width=\textwidth]{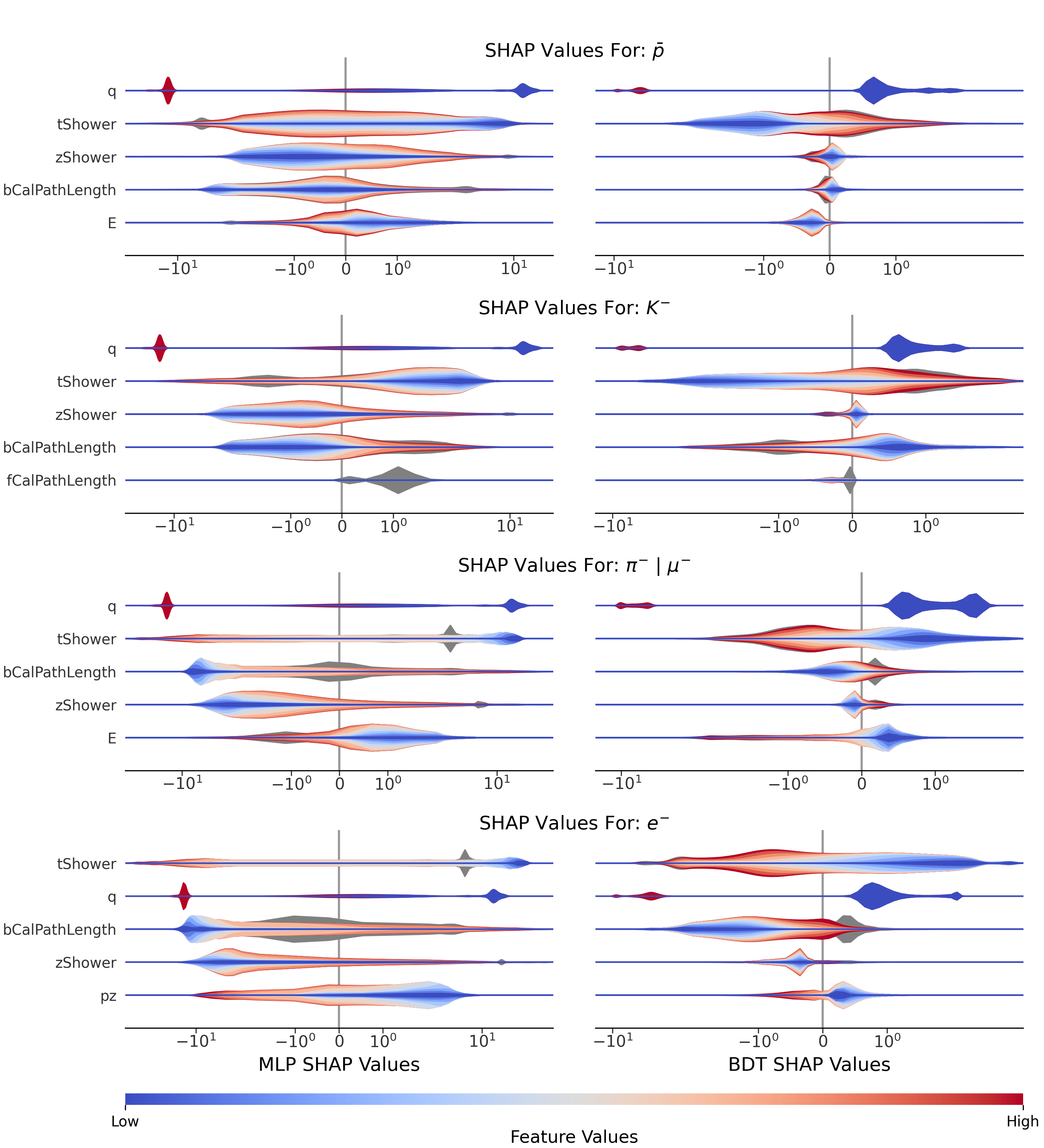}
    \caption{Layered violin plots showing SHAP value distributions for negatively charged particles. For each class, the left plot corresponds to the multilayer perceptron (MLP) classifier and the right to the boosted decision tree (BDT). Each row represents an input feature, with the horizontal axis indicating SHAP values on a symmetric logarithmic scale. The width of each violin reflects the density of SHAP values, and the internal color gradient shows the corresponding feature value, with blue representing low values and red representing high values. Gray corresponds to missing values. Features are ordered by mean absolute SHAP value averaged between both models.}
    \label{fig:negative-charged-shap}
\end{figure}

\begin{figure}[p]
    \centering
    \includegraphics[width=\textwidth]{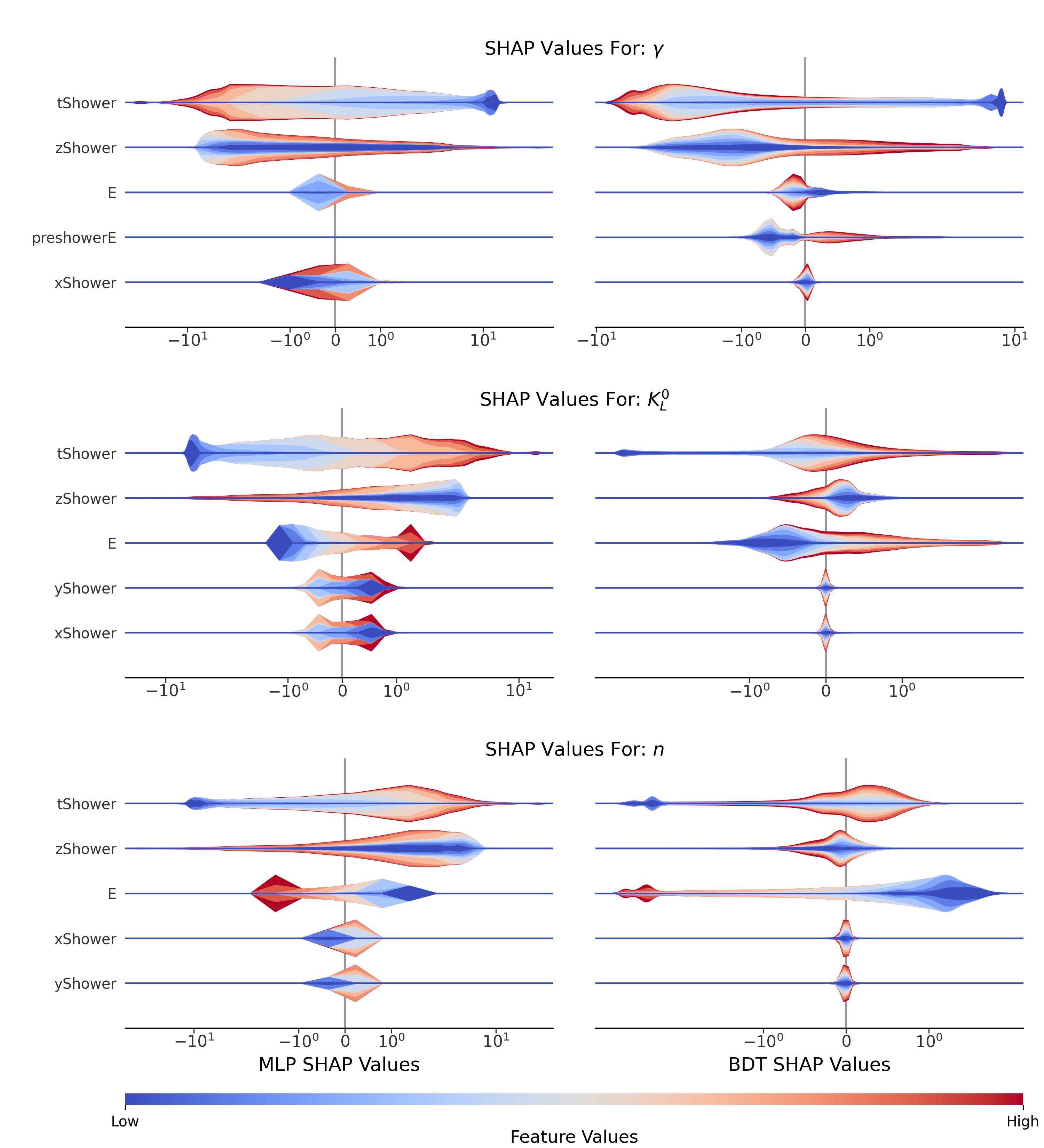}
    \caption{Layered violin plots showing SHAP value distributions for neutral particles. For each class, the left plot corresponds to the multilayer perceptron (MLP) classifier and the right to the boosted decision tree (BDT). Each row represents an input feature, with the horizontal axis indicating SHAP values on a symmetric logarithmic scale. The width of each violin reflects the density of SHAP values, and the internal color gradient shows the corresponding feature value, with blue representing low values and red representing high values. Gray corresponds to missing values. Features are ordered by mean absolute SHAP value averaged between both models. Note that the SHAP values for the preshower energy for the $\gamma$ sample for the MLP were approximately 0 for all samples, so no violin plot could be generated.}
    \label{fig:neutral-shap}
\end{figure}

To further understand the ways the ML-based PID methods are able to classify particles, SHAP values, as described in Section~\ref{subsec:shap values methods}, are calculated using the test dataset. The distributions of SHAP values for the top 5 features used to identify each positively charged, negatively charged, and neutral particle are shown in Figure~\ref{fig:positive-charged-shap}, Figure~\ref{fig:negative-charged-shap}, and Figure~\ref{fig:neutral-shap} respectively. 

For $p$, $\pi^{+}/\mu^{+}$, and $e^{+}$, as shown in Figure~\ref{fig:positive-charged-shap}, the most important feature is \texttt{tShower}. This is also the second most important feature for $K^{+}$, which has charge as its most important feature. This result is unsurprising, especially when considering the similarly important \texttt{bCalPathLength} feature, since both of these features are used in time-of-flight cuts in cuts-based PID. This is further supported by the fact that for both models, early shower times corresponded to negative SHAP values for protons and positive kaons, while early shower times corresponded to positive SHAP values for $\pi^{+}/\mu^{+}$ and $e^{+}$. This aligns with the expectation that more massive particles should have later shower times on average than less massive particles with the same momentum. Among the top features for the $p$, $K^{+}$, and $\pi^{+}/\mu^{+}$ was the energy, with low energies being associated with positive SHAP values. This makes sense because the calorimeters used by the GlueX experiment are electromagnetic calorimeters, meaning less energy is absorbed from hadronic particles. The only exception to this trend of low energy showers being associated with positive hadrons is the BDT SHAP values for the $K^{+}$, for which higher shower energy is associated with positive SHAP values. However, the SHAP values from the shower energy feature for the BDT is much smaller in magnitude on average than the corresponding magnitudes for other features used by the BDT, so this effect is likely counteracted by the larger contributions of other features. Features like \texttt{zShower} and \texttt{pz}, which show up in the top 5 features of several positive particles, may also be used for some kind of time-of-flight analysis, though the distribution of these features could also be manipulated by the event selection criteria. One such way this could occur is that kaons may decay before reaching regions of the detector with large $z$, meaning kaon events that would normally occupy the large \texttt{zShower} region of feature space may often be excluded from the data because the kaon decays before producing a shower.

Though most of the SHAP distributions for the two models are fairly similar (though often differ in magnitude, which is not inherently meaningful), the distributions for some features look extremely different. In addition to the previously mentioned disparity in the SHAP distribution of the shower energy of the $K^{+}$, the MLP is more likely to identify a particle as a $K^{+}$ if it has a moderate \texttt{tShower}, while the BDT is more likely to identify a particle as a $K^{+}$ only if it has a high \texttt{tShower}. Furthermore, while the BDT is more likely to identify a particle as a $K^{+}$ if it has a small \texttt{bCalPathLength}, the MLP does not make such a clear split between high and low \texttt{bCalPathLength}, and instead is likely making use of the feature in a more nonlinear manner.

For $\bar{p}$, $K^{-}$, $\pi^{-}/\mu^{-}$, as shown in Figure~\ref{fig:negative-charged-shap}, the most important feature is the charge, \texttt{q}. Charge is the second most important feature for $e^{-}$, which has \texttt{tShower} as its most important feature. This mirrors the cuts-based PID method, since the cuts are charge independent, then the charge is used to separate particles from antiparticles. Much like the positively charged particles, the most important features used by both models seem to be related to the time-of-flight analysis used in cuts-based PID, possibly with extra information provided by the \texttt{pz} and \texttt{zShower} features. However, unlike the positive particles, the \texttt{fCalPathLength} feature shows up as one of the most important features of the $K^{-}$, despite the feature values being largely dominated by missing values. The reason for the prominence of the \texttt{fCalPathLength} feature for the MLP classification of $K^{-}$ is not clear.

Much like the SHAP distributions for the positive particles, most of the SHAP distributions for the two models are fairly similar. However, there are still several outliers for which the SHAP distributions look very different. For example, while a higher \texttt{tShower} makes the BDT more likely to identify a particle as an $\bar{p}$ or a $K^{-}$ (which agrees with the logic of time-of-flight cuts), the opposite is true for the MLP. Similarly opposite distributions are observed for \texttt{zShower}, where the MLP is more likely to identify a particle with a large \texttt{zShower} as a $\bar{p}$ or $K^{-}$, while the BDT is less likely to identify a particle with a large \texttt{zShower} as a $\bar{p}$ or $K^{-}$. However, this effect can likely be explained by the fact that the average SHAP magnitude for the BDT is much smaller than the SHAP values for features such as \texttt{tShower}, so it is likely relying far more heavily on other features for its prediction.

For $\gamma$, $K^{0}_{L}$, and $n$, as shown in Figure~\ref{fig:neutral-shap}, the most important feature is \texttt{tShower}. Both models are more likely to classify a particle as a $\gamma$ if it has a low shower time, and more likely to identify a particle as a $K_{L}^{0}$ or $n$ if it has a high shower time. This pattern, especially when combined with other important features like \texttt{xShower}, \texttt{yShower}, and \texttt{zShower}, suggest that both models perform a modified time of flight analysis of the showers to classify neutral particles. The third most important feature for all models is the shower energy, with both models being more likely to identify a particle as a $K_{L}^{0}$ if the shower has high energy, and more likely to identify a particle as a $n$ if the shower had low energy. Most of the SHAP value distributions are similar for both models, but the distributions for $\gamma$ for \texttt{E} and \texttt{preshowerE} differ greatly. The MLP is more likely to identify particles with high shower energy as a $\gamma$, while the BDT is more likely to identify particles with low shower energy as a $\gamma$. Furthermore, while the MLP does not seem to use \texttt{preshowerE} as a factor in its predictions, the BDT learned to identify particles with large \texttt{preshowerE} as $\gamma$. This effect is likely associated with the fact that the calorimeters used by the GlueX collaboration are electromagnetic calorimeters, meaning they are more suited to absorb energy from electromagnetic showers, while hadrons may be able to travel to deeper layers of the calorimeters before producing showers, resulting in smaller \texttt{preshowerE} values.

It is clear that both models largely rely on features that are already used for cuts-based PID, though they are both able to do so with significantly higher accuracy. It is likely able to do so using extra information supplied by the momenta variables and the shower coordinates. However, it is unclear if the enhanced accuracy can be reproduced with experimental data.

\section{Conclusion} \label{sec:conclusion} 

This work compares traditional cuts-based PID methods from the GlueX experiment with multi-layered perceptron (MLP) and boosted decision tree (BDT) PID models that are trained on MC simulation data. It is found that both the MLP and BDT models outperform the standard cuts-based PID methods used by the GlueX experiment for all charged particles by an average of 18.7\% and 20.6\%, respectively. In addition to significantly better charged PID accuracies the ML methods successfully carried out neutral particle PID, which has been a long-standing challenge for the GlueX experiment. The BDT model marginally outperforms the MLP model, with an average increase of 2.1\% across all charged and neutral particles, indicating that future studies should explore the use of BDTs in future post-reconstruction PID efforts.

In addition to higher PID accuracies, the ML models are interpreted using SHAP values to study the most important features of the reconstructed events in the identification of specific particle types. Such an analysis gives insight into how the ML models make particle classifications and provides insight into which characteristics of the reconstructed events are most important in distinguishing one particle type from another. One result of interest from the model interpretation shows that for all charged particles, the distance between the vertex to BCAL is more import for ML PID than the energy deposition $(\mathrm{d}E/\mathrm{d}x)$ of a track inside of the CDC. Such insights from ML interpretation in future studies using real particle data has the potential to deepen understanding of relationships between each particle and their signature in the GlueX detector.

These results underscore the significant potential of ML to enhance PID in the analysis of GlueX experimental data. Despite the potential biases and inflated PID accuracies invoked by the simulation data set used in this study, the potential of ML to contribute to PID is evident. The use of higher energy (1 -- 12 GeV) simulation data, training with hit-based data rather than reconstructed data, and utilizing experimental data are all potential pathways to further extend the capabilities of ML for PID and bolster the success of the GlueX experiment.

\section*{Data/Software/Code Availability}

All the software and data used in this research is available in this public \href{https://doi.org/10.5281/zenodo.16878964}{Zenodo repository}.



\bibliographystyle{JHEP}
\bibliography{biblio}

\end{document}